\documentclass[aps,preprint]{revtex4}%
\usepackage{amsfonts}
\usepackage{amsmath}
\usepackage{amssymb}
\usepackage{graphicx}%
\begin{document}
\title[Relativistic Conservation Laws]{Illustrating Some Implications of the Conservation Laws in Relativistic Mechanics}
\author{Timothy H. Boyer}
\affiliation{Department of Physics, City College of the City University of New York, New
York, New York 10031}

\begin{abstract}
The conservation laws of nonrelativistic and relativistic systems are reviewed
and some simple illustrations are provided for the restrictive nature of the
relativistic conservation law involving the center of energy compared to the
nonrelativistic conservation law for the center of restmass. \ Extension of
the nonrelativistic interaction of particles through a potential to a system
which is Lorentz-invariant through order $v^{2}/c^{2}$ is found to require new
velocity- and acceleration-dependent forces which are suggestive of field
theory where the no-interaction theorem of Currie, Jordan, and Sudershan does
not hold. \ 

\end{abstract}
\maketitle

\section{Introduction}

At the beginning of the twentieth century, the mismatch between mechanics and
electromagnetism led to the creation of two new theories, quantum mechanics
and special relativity. \ Today the mismatch between mechanics and relativity
is still not appreciated by many students of physics. \ There is a tendency to
believe that one can pass from a nonrelativistic mechanical theory over to a
relativistic theory simply by using relativistic expressions for the
mechanical energy and momentum of a particle while retaining a general
nonrelativistic potential.\cite{belief} \ In 1963 Currie, Jordan, and
Sudarshan\cite{CJSn} proved their "no-interaction theorem," pointing out just
how wrong this point of view really is. \ They proved that two point particles
which satisfy the conservation laws of Lorentz-invariant mechanics simply can
not interact except through point contact forces. \ This no-interaction
theorem is sometimes alluded to in mechanics texts as an afterthought in
discussions of one-particle relativistic motion.\cite{Goldstein} \ However,
the text books seem to contain no physical examples of what is involved in the
theorem. \ In the present article we wish to rectify this omission by
reviewing the conservation laws associated with Lorentz invariance and then
exploring some simple mechanical examples which suggest the need for a
relativistic field theory such as classical electrodynamics.

We start by listing the conservation laws of physical systems when in the
presence of external forces. \ These laws are associated with momentum,
energy, angular momentum, and center-of-energy motion. \ It is the fourth law
associated with the symmetry of systems under change of inertial frame which
is so different between nonrelativistic and relativistic systems. \ We
emphasize that this fourth conservation law is linked to the continuous flow
of energy in relativistic systems, and this continuous flow of energy places
severe restrictions on allowed systems. \ In the nonrelativistic limit, this
fourth conservation law reduces to the continuous flow of rest mass, so that
the fourth conservation law merely restates the law of momentum conservation
for particles, and places no restrictions upon nonrelativistic mechanical
systems. \ 

As our first example, we consider an external force applied to a single
particle and use the conservation laws to derive both the Galilean- and the
Lorentz-invariant expressions for mechanical energy and momentum. \ Second, we
consider the collision of two point particles at a single point, and show that
all the conservation laws of either Galilean- or Lorentz-invariant physics can
be satisfied by such single-point collisions. \ If one considers the
collisions of one particle which is described by Galilean-invariant mechanical
expressions and one particle which is described by Lorentz-invariant
mechanical expressions, then the conservation laws of momentum, energy, and
angular momentum can still be satisfied in a fixed inertial frame, but the
outcome of the collision will depend specifically on the choice of inertial
frame; the fourth conservation law associated with Galilean or Lorentz
invariance will not hold in any inertial frame. \ The third set of examples
involves collisions of particles which do not interact at a single point but
through a potential. \ We show that the continuous flow of energy does not
hold if the particles interact through a rigid rod or through a general
potential. \ The one case where the Lorentz-invariant conservation law can be
found to hold involves a constant force between the particles independent of
their separation, a situation which can be reinterpreted within familiar
physics as the collision of two electrically charged capacitor plates. \ The
fourth section discusses the modification of a general potential
$V(|\mathbf{r}_{1}-\mathbf{r}_{2}|)$ by adding velocity-dependent terms so as
to make the interaction of particles Lorentz-invariant through order
$v^{2}/c^{2}.$ \ Such a modification leads to velocity- and
acceleration-dependent forces between the particles and suggests a
field-theory interaction for Lorentz-invariant behavior. \ The Darwin
Lagrangian is seen to follow as the approximately Lorentz-invariant extension
from the Coulomb potential, and the Darwin Lagrangian is known to be the valid
$v^{2}/c^{2}$ approximation to classical electrodynamics. \ Finally we present
a closing summary and discussion.

\section{The Conservation Laws in Relativistic and Nonrelativistic Physics}

In the usual formulations of mechanics and field theory, the conservation laws
of physical systems are associated with the generators of symmetry
transformations for the systems. \ These include linear momentum conservation
associated with space-translation invariance, energy conservation associated
with time-translation invariance, angular momentum conservation associated
with rotation invariance, and finally two distinct conservation laws
associated with invariance under transformation to a new inertial frame. \ The
generator of Galilean transformations is the total system restmass times the
center of mass, while the generator of Lorentz transformations is the total
energy times the center of energy.\cite{CV} \ 

In the presence of external forces $\mathbf{F}_{ext\;i}$ on the particles
$m_{i}$ located at positions $\mathbf{r}_{i}$ moving with velocity
$\mathbf{v}_{i},$ the conservation laws for a Lorentz-invariant mechanical
system or field theory take the following forms.\cite{CofE} \ The sum of the
external forces on the system gives the time rate of change of system momentum
$\mathbf{P}$%
\begin{equation}%
{\displaystyle\sum\limits_{i}}
\mathbf{F}_{ext\;i}=\frac{d\mathbf{P}}{dt}%
\end{equation}
The total power delivered to the system by the external forces gives the time
rate of change of system energy $U$%
\begin{equation}%
{\displaystyle\sum\limits_{i}}
\mathbf{F}_{ext\;i}\cdot\mathbf{v}_{i}=\frac{dU}{dt}%
\end{equation}
The sum of the external torques gives the time rate of change of system
angular momentum $\mathbf{L}$%
\begin{equation}%
{\displaystyle\sum\limits_{i}}
\mathbf{r}_{i}\times\mathbf{F}_{ext\;i}=\frac{d\mathbf{L}}{dt}%
\end{equation}
The external-power-weighted position equals the time rate of change of the
quantity (the total energy $U$ times the center of energy $\overrightarrow
{\mathcal{X}}_{energy})$ minus $c^{2}$ times the system momentum%
\begin{equation}%
{\displaystyle\sum\limits_{i}}
\left(  \mathbf{F}_{ext\;i}\cdot\mathbf{v}_{i}\right)  \mathbf{r}_{i}=\frac
{d}{dt}(U\overrightarrow{\mathcal{X}}_{energy})-c^{2}\mathbf{P}%
\end{equation}
Here the center of energy $\overrightarrow{\mathcal{X}}_{energy}$ is defined
so that%
\begin{equation}
U\overrightarrow{\mathcal{X}}_{energy}=%
{\displaystyle\sum\limits_{i}}
U_{i}\mathbf{r}_{i}+%
{\displaystyle\int}
d^{3}ru(\mathbf{r)r}%
\end{equation}
where $U_{i}$ is the mechanical energy of the $i$th particle and
$u(\mathbf{r)}$ is the continuous system energy density at position
$\mathbf{r}$. \ This last conservation law (4) expresses the continuous flow
of energy in Lorentz-invariant systems. \ In an isolated system where no
external forces are present, the linear momentum, energy, and angular momentum
are all constants in time, and the center of energy $\overrightarrow
{\mathcal{X}}_{energy}$ moves with constant velocity $d\overrightarrow
{\mathcal{X}}_{energy}/dt=c^{2}\mathbf{P}/U$ since the energy $U$ and momentum
$\mathbf{P}$ in Eq. (4) are both constant.

The conservation laws for Galilean invariance involve exactly the same
expression in Eqs. (1) -(3) concerning linear momentum, energy, and angular
momentum. \ However the last equation (4) involving the center of energy takes
the degenerate form%
\begin{equation}
0=\frac{d}{dt}\left(  (%
{\displaystyle\sum\limits_{i}}
m_{i})\overrightarrow{\mathcal{X}}_{mass}\right)  -\mathbf{P}%
\end{equation}
obtained by dividing Eq. (4) through by $c^{2}$ and taking the $c\rightarrow
\infty$ limit. \ In this case, the particle energy $U_{i}$ divided by $c^{2}$
becomes the rest mass, $U_{i}/c^{2}\rightarrow m_{i}$ when we take the limit
$c\rightarrow\infty$. \ We note that in this limit, $(\mathbf{F}_{ext\;i}%
\cdot\mathbf{v}_{i})\mathbf{r}_{i}/c^{2}\rightarrow0$ so that the external
forces do not enter this fourth (and last) Galilean conservation law. \ Within
Galilean invariance, the fourth conservation law expresses the continuous flow
of restmass and this continuity is not interrupted by the presence of external
forces which may introduce linear momentum, energy, and angular momentum into
the system.

\section{Derivations of Expressions for Mechanical Linear Momentum and Energy
of a Particle}

We can use these conservation laws when applied to a single particle to derive
the nonrelativistic and relativistic expressions for mechanical energy and
momentum. \ If a single particle of mass $m$ experience a force $\mathbf{F}$,
then according to the first conservation law in Eq. (1), the particle
mechanical momentum $\mathbf{p}$ (which is the entire momentum of the system)
changes as%
\begin{equation}
\mathbf{F=}\frac{d\mathbf{p}}{dt}%
\end{equation}
while the change in mechanical energy $U$ of the particle is given by the
second conservation law in Eq. (2)
\begin{equation}
\mathbf{F\cdot}\frac{d\mathbf{r}}{dt}=\frac{dU}{dt}%
\end{equation}
The third conservation law is not needed here for the present interest of
exploring the linear momentum and energy of a point particle. \ The fourth
conservation law takes a different form for Galilean and Lorentz invariance.

\subsection{Galilean Invariance}

The center of mass of a one-particle system is clearly located at the position
$\mathbf{r}$ of the single particle. \ Then the fourth Galilean conservation
law (6) for the center of mass of this one-particle system satisfies
\begin{equation}
\frac{d}{dt}(m\mathbf{r)=p}%
\end{equation}
so that the mechanical momentum of the nonrelativistic particle is identified
in terms of the time rate of change of the mass times the position of the
particle. \ Then since the mass $m$ is a constant in time, the particle
mechanical momentum must be given by the familiar nonrelativistic expression%
\begin{equation}
m\frac{d\mathbf{r}}{dt}=\mathbf{p}%
\end{equation}
Then from Eqs. (7) and (8), $dU/dt=(d\mathbf{p/}dt)\cdot(d\mathbf{r}/dt),$ and
then from Eq. (10) it follows that $U=\frac{1}{2}m\left(  d\mathbf{r}%
/dt\right)  ^{2}+const$ where the constant corresponds to an undetermined zero
of energy. \ Usually for nonrelativistic mechanical energy, the constant is
chosen to vanish, giving the familiar
\begin{equation}
U=(1/2)m(d\mathbf{r}/dt)^{2}%
\end{equation}
of nonrelativistic physics. \ 

\subsection{Lorentz Invariance}

On the other hand, according to Lorentz invariance, the conservation law (4)
for the center of energy of this one-particle system becomes%

\begin{equation}
\left(  \mathbf{F\cdot}\frac{d\mathbf{r}}{dt}\right)  \mathbf{r=}\frac{d}%
{dt}\left(  U\mathbf{r}\right)  -c^{2}\mathbf{p}%
\end{equation}
Expanding the time derivative $d(U\mathbf{r})/dt=(dU/dt)\mathbf{r+}%
U(d\mathbf{r}/dt)$ in Eq (12), noting the power relation $\mathbf{F}%
\cdot(d\mathbf{r}/dt)=dU/dt$ in Eq. (8), and cancelling two terms, we have%
\begin{equation}
0=U\frac{d\mathbf{r}}{dt}-c^{2}\mathbf{p}%
\end{equation}
However, we can then eliminate $d\mathbf{r/}dt$ between $dU/dt=(d\mathbf{p}%
/dt)\cdot(d\mathbf{r}/dt)$ from Eqs. (7) and (8) and $U(d\mathbf{r}%
/dt)=c^{2}\mathbf{p}$ in Eq. (13) to obtain%
\begin{equation}
\frac{dU}{dt}=\frac{d\mathbf{p}}{dt}\cdot\frac{c^{2}\mathbf{p}}{U}%
\end{equation}
which has the solution $U^{2}=c^{2}p^{2}+const.$ \ If we choose the constant
as $m^{2}c^{4},$ then we have the familiar relativistic expression for
particle mechanical energy
\begin{equation}
U=(c^{2}p^{2}+m^{2}c^{4})^{1/2}%
\end{equation}
\ where $m$ is the particle restmass. \ Also, \ we can use Eq. (15) to
eliminate the energy $U$ in Eq. (13) and solve for $\mathbf{p}$ to obtain the
familiar relativistic expression for the linear momentum
\begin{equation}
\mathbf{p=}m\gamma(d\mathbf{r/}dt)
\end{equation}
where $\gamma=(1-v^{2}/c^{2})^{-1/2}$ with $\mathbf{v}=d\mathbf{r}/dt$ and
$v=|\mathbf{v}|.$

\section{Relativistic Mechanics of Particle Collisions}

\subsection{Familiar Particle Collisions at Points}

Particle collisions at points are familiar textbook subjects in both
nonrelativistic and relativistic mechanics. \ In these cases, there are no
external forces and the particle momenta and energies correspond to the
mechanical energy and momentum given by Eqs. (10) and (11) or by Eqs. (15) and
(16) for the nonrelativistic and relativistic cases respectively. \ The
collisions are assumed to conserve total energy and total momentum. \ Since
the particle collisions occur at a single point $\mathbf{r}$, the conservation
laws of energy and momentum are sufficient to guarantee the additional laws of
angular momentum conservation, and constant motion of the center of mass in
the Galilean case or constant motion of the center of (mechanical) energy in
the Lorentz case. \ In these cases, the conservation laws can be applied in
any inertial frame and will be found to hold in any other inertial frame. \ 

\subsubsection{Relativistic Collision}

Here we introduce an illustration of elastic collisions at a point; later we
will reinvestigate the collision when interaction potentials are introduced.
\ For simplicity, we consider two particles $m_{1}$ and $m_{2}$ of equal mass
$m=m_{1}=m_{2}$, the first of which is approaching with speed $v$ along the
negative $x$-axis, $x_{1}=vt$ for $t<0$ and the second of which is initially
at rest at the coordinate origin, $x_{2}=0$ for $t<0.$ \ Thus before the
collision, the system energy times the center of energy is given by $(m\gamma
c^{2}+mc^{2})\mathcal{X}_{energy}=(m\gamma c^{2})vt+mc^{2}0$ so that the
center of energy moves with constant velocity%
\begin{equation}
\mathcal{X}_{energy}=\frac{(m\gamma c^{2})vt}{(m\gamma c^{2}+mc^{2})}%
\end{equation}
where $\gamma=(1-v^{2}/c^{2})^{-1/2}.$ \ We assume that the two particles have
no associated potential energy between them so that the particles will not
interact until the point collision at the origin at time $t=0$. \ At this
collision, the particles exchange energy and momentum. \ After the collision,
the first particle comes to rest at the origin while the second particle
carries the energy $m\gamma c^{2}$ along the positive $x$-axis. The center of
energy is given by $(mc^{2}+m\gamma c^{2})\mathcal{X}_{energy}=mc^{2}%
0+(m\gamma c^{2})vt,$ so that the motion is again given by Eq. (17). \ In this
collision at a single point, the system center of energy moves with constant
velocity, and the fourth relativistic conservation law (4) is indeed
satisfied. \ 

\subsubsection{Nonrelativistic Collision}

We could also treat this collision problem in nonrelativistic physics. \ In
this case the fourth conservation law involves the center of restmass
following from $(m+m)\mathcal{X}_{mass}=mvt+m0,$%
\begin{equation}
\mathcal{X}_{mass}=\frac{mvt}{2m}%
\end{equation}
After the collision at the origin, the center or restmass is given by
$(m+m)\mathcal{X}_{mass}=m0+mvt,$ which again leads to Eq. (18). \ Thus the
center of restmass moves with constant velocity when nonrelativistic
expressions are used for the mechanical energy and momentum.

\subsubsection{Mixed Relativistic-Nonrelativistic Collision}

It is also possible to consider the elastic collisions of point particles even
when one colliding particle is described by nonrelativistic energy and
momentum and the other by relativistic energy and momentum. \ Within a single
inertial frame, the momentum, energy, and angular momentum conservation laws
can all be handled satisfactorily. \ Thus, for our simple example, we can
describe the energy and momentum of the first (incoming) particle by
relativistic expressions $U_{1}=m\gamma c^{2}$ and $p_{1}=m\gamma v$ while
using nonrelativistic expressions for the second particle. \ In this case, the
first particle would not be brought to rest on collision with the particle of
equal mass which had been sitting at the origin. \ Rather, we would solve for
the final velocities of both particles using the momentum and energy
conservation laws%
\begin{equation}
m\gamma v+0=m\gamma_{1}v_{1}+mv_{2}%
\end{equation}
and%
\begin{equation}
m\gamma c^{2}+0=m\gamma_{1}c^{2}+\frac{1}{2}mv_{2}^{2}%
\end{equation}
where $\gamma=(1-v^{2}/c^{2})^{-1/2}$ and\ $\gamma_{1}=(1-v_{1}^{2}%
/c^{2})^{-1/2}.$\ \ We have two equations in the two unknown final velocities
$v_{1}$ and $v_{2},$ and so can solve for these quantities in terms of the
initial incoming particle velocity $v.$\cite{mixbox} \ Our example will
satisfy neither constant motion of the center of energy nor constant motion of
the center of restmass. \ Such a situation of mixed relativistic and
nonrelativistic expressions involves neither Galilean invariance nor Lorentz
invariance, and the collision outcome will depend upon the specific inertial
frame in which the conservation laws for the collision are applied. \ Since
most physicists are fully aware of only the first three conservation laws, the
failure of the fourth conservation law on mixing Galilean- and
Lorentz-invariant systems (usually nonrelativistic mechanics combined with
electrodynamics) is rarely noted.\cite{Rohrlich}

\subsection{Particle Collisions Involving Potentials}

Although point collisions within nonrelativistic or relativistic mechanics are
quite satisfactory, the introduction of potentials between relativistic
particles is quite another matter. \ Indeed, the no-interaction theorem of
Currie, Jordan, and Sudarshan\cite{CJSn} states that within relativistic
mechanics the use of such potentials is completely forbidden by our usual
ideas of mechanics. \ The Lorentz-invariant center-of-energy requirement of
Eq. (4) is so restrictive that it does not allow mechanical interactions
through potentials. \ In order to escape the no-interaction theorem, one must
turn to a field theory.

\subsubsection{Collision Through a Rigid Pole}

Here we illustrate what is involved in the no-interaction theorem by
reconsidering the collision of our two particles. \ Let us now consider the
same collision of two relativistic particles but this time the first particle
(at $x_{1}=vt$ for early times) carries a massless rigid pole of length $R$
which precedes the particle. \ This rigid pole produces a collision with the
second particle when there is still a separation $R$ between them. \ Now the
collision occurs at time $t=-R/v.$ The first particle stops at $x_{1}=-R$
while its energy and momentum are transferred instantaneously to the second
particle which moves as $x_{2}=v(t+R/v).$ \ In this case, the system center of
energy does not move with constant velocity, but rather makes a discontinuous
jump as the energy is transferred instantaneously along the pole to the other
particle. \ Before the collision, the center of energy motion is given by Eq.
(17), but after the collision, the center of energy motion follows from
$(mc^{2}+m\gamma c^{2})\mathcal{X}_{energy}=mc^{2}(-R)+m\gamma c^{2}v(t+R/v)$
giving
\begin{equation}
\mathcal{X}_{energy}=\frac{(m\gamma c^{2})vt+mc^{2}(\gamma-1)R}{(mc^{2}%
+m\gamma c^{2})}%
\end{equation}
which does not agree with Eq. (17).

Since such rigid-pole collisions do not satisfy the conservation laws
associated with Lorentz-invariant behavior, they are not allowed in
relativistic theory. \ Indeed in kinematic discussions of special relativity,
students are regularly warned against the possibility of rigid poles.
\ However, we should emphasize that such rigid poles do allow the
much-less-stringent condition (6) of Galilean invariance which corresponds to
the continuous flow of rest mass. \ We saw above that if we treat the
two-particle collision using nonrelativistic physics, then the center of
restmass $\mathcal{X}_{mass}$ moves with the constant velocity of Eq. (18)
before the collision. \ However, the center of restmass also moves with
constant velocity even after the two nonrelativistic particles collide through
a rigid pole. \ Thus after the collision, the first particle comes to rest at
$-R$ while the second particle moves off from the origin with a position given
by $x_{2}=v(t+R/v)$ since the collision occurred at time $t=-R/v.$ \ Thus the
center of restmass is calculated as $(m+m)\mathcal{X}_{mass}%
=m(-R)+m[v(t+R/v)]$ which gives exactly the same result as in Eq. (18) which
held before the collision. \ The rigid pole transfers energy, not restmass,
and so gives a discontinuous jump in the center of energy but not in the
center of restmass.

\subsubsection{Collision Through a General Potential $V(|x_{2}-x_{1}|)$}

The use of a rigid pole in the collision of our example is equivalent to
considering a hard-sphere potential between the particles. \ Our example
illustrates that a hard-sphere potential leads to an instantaneous transfer of
energy and hence to a failure of the Lorentz-invariant conservation law
requiring a constant velocity for the center of energy in an isolated system.
\ The use of a hard-sphere potential is an extreme situation for a potential.
\ One might hope that by use of a softer interaction (perhaps through massless
springs) one might be able to accommodate potentials into relativistic
mechanics. \ However, this can not be done in general. \ The same basic flaw
keeps reappearing; there is a transfer of energy between spatial points
without a continuous energy flow through the intervening space. \ 

We can illustrate this difficulty by considering the\ collision along the
$x$-axis of two particles $m_{1}$ and $m_{2}$ of relativistic energy and
momentum interacting through a general potential $V(|x_{2}-x_{1}|)$ dependent
upon the separation between the particles where we assign the center of energy
associated with the potential energy to a location half-way between the
particles. \ In this case the energy times the center of energy takes the form%
\begin{equation}
U\mathcal{X}_{energy}=m_{1}\gamma_{1}c^{2}x_{1}+m_{2}\gamma_{2}c^{2}%
x_{2}+V(|x_{2}-x_{1}|)\frac{(x_{1}+x_{2})}{2}%
\end{equation}
We require that the particles experience forces associated with Newton's
second law arising from the potential $V(|x_{2}-x_{1}|)$ and for simplicity
assume $x_{1}\,<x_{2}$ so that $|x_{2}-x_{1}|=x_{2}-x_{1}.$ \ Then the
expressions (7) and (8) for Newton's second law and particle mechanical energy
change take the form
\begin{equation}
\frac{dp_{1}}{dt}=V^{\prime}(|x_{2}-x_{1}|)\text{ \ \ and \ \ }\frac{dp_{2}%
}{dt}=-V^{\prime}(|x_{2}-x_{1}|)
\end{equation}
and%
\begin{equation}
\frac{d}{dt}(m_{1}\gamma_{1}c^{2})=\frac{dp_{1}}{dt}v_{1}\text{ \ \ and
\ \ \ }\frac{d}{dt}(m_{2}\gamma_{2}c^{2})=\frac{dp_{2}}{dt}v_{2}%
\end{equation}
where $V^{\prime}(|x_{2}-x_{1}|)$ refers to the derivative of the potential
with respect to its argument. Then the total energy $U=m_{1}\gamma_{1}%
c^{2}+m_{2}\gamma_{2}c^{2}+V(|x_{2}-x_{1}|)$ and momentum $P=p_{1}+p_{2}$ are
both constant in time and%
\begin{align}
\frac{d}{dt}(U\mathcal{X}_{energy}) &  =U\frac{d\mathcal{X}_{energy}}%
{dt}=\frac{d}{dt}\left(  m_{1}\gamma_{1}c^{2}x_{1}+m_{2}\gamma_{2}c^{2}%
x_{2}+V(|x_{2}-x_{1}|)\frac{(x_{1}+x_{2})}{2}\right)  \nonumber\\
&  =\frac{d}{dt}(m_{1}\gamma_{1}c^{2})x_{1}+\frac{d}{dt}(m_{2}\gamma_{2}%
c^{2})x_{2}+V^{\prime}(|x_{2}-x_{1}|)(v_{2}-v_{1})\frac{(x_{1}+x_{2})}%
{2}\nonumber\\
&  +m_{1}\gamma_{1}c^{2}v_{1}+m_{2}\gamma_{2}c^{2}v_{2}+V(|x_{2}-x_{1}%
|)\frac{(v_{1}+v_{2})}{2}\nonumber\\
&  =\left(  \frac{dp_{1}}{dt}-V^{\prime}(|x_{2}-x_{1}|)\right)  v_{1}%
x_{1}+\left(  \frac{dp_{2}}{dt}-V^{\prime}(|x_{2}-x_{1}|)\right)  v_{2}%
x_{2}+c^{2}(p_{1}+p_{2})\nonumber\\
&  +V(|x_{2}-x_{1}|)\frac{(v_{1}+v_{2})}{2}+\frac{1}{2}V^{\prime}(|x_{2}%
-x_{1}|)(v_{1}+v_{2})(x_{1}-x_{2})\nonumber\\
&  =c^{2}P+[V(|x_{2}-x_{1}|)-V^{\prime}(|x_{2}-x_{1}|)(x_{2}-x_{1}%
)]\frac{(v_{1}+v_{2})}{2}%
\end{align}
where we have used the equations of motion to eliminate two terms in round
brackets. \ Since $x_{1},$ $x_{2},$ $v_{1},$ and $v_{2}$ can be chosen
arbitrarily, the only way to satisfy the fourth Lorentz-invariant conservation
law is to have the square parenthesis in the last line of Eq. (25) vanish.
\ This requires that
\begin{equation}
V(|x_{2}-x_{1}|)=k|x_{2}-x_{1}|
\end{equation}
where $k$ is a constant, and corresponds to a constant force between the
particles. \ In this case, there is a smooth rather than a sudden transfer of
energy between the particles. \ However, this unique potential giving
Lorentz-invariant behavior for the center of energy seems surprising as an
interparticle potential since it increases in magnitude with distance rather
than decreases. \ There is no asymptotic region where the particles can be
regarded as unaffected by the other particle. \ Actually, this unique
potential is quite familiar, not as the potential between two particles but
rather as the electrostatic potential energy for two uniformly charged
parallel plates of large area $A=L\times L$ and small separation $|x_{2}%
-x_{1}|<<L.$ \ In this case, the uniform electrostatic field between the
plates indeed contributes an energy whose natural center of energy is halfway
between the plates. \ Here the transfer of energy between the field and the
plate is carried out locally at each plate, with the field energy either
appearing or disappearing as the plate mechanical energy decreases or
increases. \ This parallel-plate example has been discusses in more detail in
two other analyses.\cite{parallel} \ The \textit{interparticle} potential
which leads to the potential (24) between the (\textit{multiparticle})
parallel plates is the Coulomb potential between charged particles, which is
part of classical electrodynamic field theory and is not part of a
Lorentz-invariant mechanical system of particles with potentials. \ The
no-interaction theorem of Currie, Jordan, and Sudarshan\cite{CJSn} applies to
Lorentz-invariant mechanical systems and does not apply to field theories.

\section{Nonrelativistic Lagrangians for Particles and Lorentz-Invariant
Extension to Order $v^{2}/c^{2}$}

Although the no-interaction theorem of relativistic mechanics indicates that
it is hopeless to create a conventional fully relativistic mechanical theory
with an arbitrary potential, one might try to extend nonrelativistic mechanics
toward a theory which is approximately relativistic to higher orders in $v/c,$
where $v$ is the typical particle velocity in some inertial frame. \ This
would give us some indication of just when we need to turn to field theory in
order to obtain a fully relativistic system.

The nonrelativistic system of two particles interacting through a potential
$V(|\mathbf{r}_{1}-\mathbf{r}_{2}|)$ can be described by a Lagrangian
\begin{equation}
L(\mathbf{r}_{1},\mathbf{r}_{2},\mathbf{\dot{r}}_{1},\mathbf{\dot{r}}%
_{2})=\frac{1}{2}m_{1}\mathbf{\dot{r}}_{1}^{2}+\frac{1}{2}m_{2}\mathbf{\dot
{r}}_{2}^{2}-V(|\mathbf{r}_{1}-\mathbf{r}_{2}|)
\end{equation}
The invariance of this Lagrangian under spacetime translations and spatial
rotations leads to the conservation laws for linear momentum, energy, and
angular momentum. \ In order to extend this system toward an approximately
Lorentz-invariant system, we must preserve the invariance of the Lagrangian
under spacetime translations \ and spatial rotations while adding (small)
additional space-dependent and velocity-dependent functions of order
$v^{2}/c^{2}$.

\subsection{Lagrangian Lorentz-Invariant Through Order $v^{2}/c^{2}$}

Since the energy and momentum of an isolated system should form a Lorentz
four-vector, we expect the potential energy $V(|\mathbf{r}_{1}-\mathbf{r}%
_{2}|)$ to be related to momentum in a different inertial frame. Thus in a
moving frame, we expect to find velocity-dependent forces between the
particles in addition to the position-dependent forces found in the original
frame$.$ \ If we require Lorentz invariance through order $v^{2}/c^{2},$ then
the velocity-dependent terms must appear in the Lagrangian in any inertial
frame. \ By working backwards from the requirement of Lorentz invariance given
in Eq. (4) and requiring that the condition hold through order $v^{2}/c^{2},$
we find that the Lagrangian extended from the nonrelativistic expression can
be written as\cite{Essen}%
\begin{align}
L(\mathbf{r}_{1},\mathbf{r}_{2},\mathbf{\dot{r}}_{1},\mathbf{\dot{r}}_{2})  &
=-m_{1}c^{2}(1-\mathbf{\dot{r}}_{1}^{2}/c^{2})^{1/2}-m_{2}c^{2}(1-\mathbf{\dot
{r}}_{2}^{2}/c^{2})^{1/2}-V(|\mathbf{r}_{1}-\mathbf{r}_{2}|)\nonumber\\
&  +\frac{1}{2}V(|\mathbf{r}_{1}-\mathbf{r}_{2}|)\frac{\mathbf{\dot{r}}%
_{1}\cdot\mathbf{\dot{r}}_{2}}{c^{2}}-\frac{1}{2}V^{\prime}(|\mathbf{r}%
_{1}-\mathbf{r}_{2}|)\frac{\mathbf{\dot{r}}_{1}\cdot(\mathbf{r}_{1}%
-\mathbf{r}_{2})\mathbf{\dot{r}}_{2}\cdot(\mathbf{r}_{1}-\mathbf{r}_{2}%
)}{c^{2}\left\vert \mathbf{r}_{1}-\mathbf{r}_{2}\right\vert }%
\end{align}
where we have introduced the Lagrangian terms leading to relativistic
mechanical momentum and energy for the particles, and where $V^{\prime
}(|\mathbf{r}_{1}-\mathbf{r}_{2}|)$ refers to the derivative of the potential
function with respect to its argument. \ 

We can check the Lorentz invariance of this Lagrangian through order
$v^{2}/c^{2}$ by showing that Eq. (4) (with $\mathbf{F}_{ext\;i}=0)$ holds
through this order; in other words, the system center of energy moves with
constant velocity through order $v^{2}/c^{2}.$ \ The system energy $U$ times
the center of energy of the system $\overrightarrow{X}$ through zero-order in
$v/c$ is given by%
\begin{equation}
U\overrightarrow{X}=m_{1}(c^{2}+\frac{1}{2}\mathbf{\dot{r}}_{1}^{2}%
)\mathbf{r}_{1}+m_{2}(c^{2}+\frac{1}{2}\mathbf{\dot{r}}_{2}^{2})\mathbf{r}%
_{2}+V(|\mathbf{r}_{1}-\mathbf{r}_{2}|)\frac{(\mathbf{r}_{1}+\mathbf{r}_{2}%
)}{2}%
\end{equation}
corresponding to the restmass energy and kinetic energy of the two particles
located at their respective positions $\mathbf{r}_{1}$ and $\mathbf{r}_{2}$
plus the interaction potential energy located half way between the positions
of the two particles. \ Since the Lagrangian in Eq. (28) has no explicit time
dependence, the system energy $U$ is constant in time. \ Taking the time
derivative of Eq. (29), we find%
\begin{align}
\frac{d}{dt}(U\overrightarrow{X})  &  =U\frac{d\overrightarrow{X}}{dt}%
=m_{1}(c^{2}+\frac{1}{2}\mathbf{\dot{r}}_{1}^{2})\mathbf{\dot{r}}_{1}%
+m_{2}(c^{2}+\frac{1}{2}\mathbf{\dot{r}}_{2}^{2})\mathbf{\dot{r}}_{2}%
+(m_{1}\mathbf{\ddot{r}}_{1}\mathbf{\cdot\dot{r}}_{1})\mathbf{r}_{1}%
+(m_{2}\mathbf{\ddot{r}}_{2}\mathbf{\cdot\dot{r}}_{2})\mathbf{r}%
_{2}\nonumber\\
&  +\frac{1}{2}V(|\mathbf{r}_{1}-\mathbf{r}_{2}|)(\mathbf{\dot{r}}%
_{1}\mathbf{+\dot{r}}_{2}\mathbf{)+}\frac{1}{2}V^{\prime}(|\mathbf{r}%
_{1}-\mathbf{r}_{2}|)\frac{(\mathbf{\dot{r}}_{1}-\mathbf{\dot{r}}_{2}%
)}{|\mathbf{r}_{1}-\mathbf{r}_{2}|}\cdot(\mathbf{r}_{1}-\mathbf{r}%
_{2})(\mathbf{r}_{1}+\mathbf{r}_{2})
\end{align}
It is sufficient to use the nonrelativistic equations of motion,%
\begin{equation}
m_{1}\mathbf{\ddot{r}}_{1}=-V^{\prime}(|\mathbf{r}_{1}-\mathbf{r}_{2}%
|)\frac{(\mathbf{r}_{1}-\mathbf{r}_{2})}{\left\vert \mathbf{r}_{1}%
-\mathbf{r}_{2}\right\vert }%
\end{equation}%
\begin{equation}
m_{2}\mathbf{\ddot{r}}_{2}=V^{\prime}(|\mathbf{r}_{1}-\mathbf{r}_{2}%
|)\frac{(\mathbf{r}_{1}-\mathbf{r}_{2})}{\left\vert \mathbf{r}_{1}%
-\mathbf{r}_{2}\right\vert }%
\end{equation}
to transform Eq. (30) into the form%
\begin{align}
\frac{d}{dt}(U\overrightarrow{X})  &  =U\frac{d\overrightarrow{X}}{dt}%
=m_{1}(c^{2}+\frac{1}{2}\mathbf{\dot{r}}_{1}^{2})\mathbf{\dot{r}}_{1}%
+m_{2}(c^{2}+\frac{1}{2}\mathbf{\dot{r}}_{2}^{2})\mathbf{\dot{r}}%
_{2}\nonumber\\
&  -\left(  V^{\prime}(|\mathbf{r}_{1}-\mathbf{r}_{2}|)\frac{(\mathbf{r}%
_{1}-\mathbf{r}_{2})}{\left\vert \mathbf{r}_{1}-\mathbf{r}_{2}\right\vert
}\mathbf{\cdot\dot{r}}_{1}\right)  \mathbf{r}_{1}+\left(  V^{\prime
}(|\mathbf{r}_{1}-\mathbf{r}_{2}|)\frac{(\mathbf{r}_{1}-\mathbf{r}_{2}%
)}{\left\vert \mathbf{r}_{1}-\mathbf{r}_{2}\right\vert }\mathbf{\cdot\dot{r}%
}_{2}\right)  \mathbf{r}_{2}\nonumber\\
&  +\frac{1}{2}V(|\mathbf{r}_{1}-\mathbf{r}_{2}|)(\mathbf{\dot{r}}%
_{1}\mathbf{+\dot{r}}_{2}\mathbf{)+}\frac{1}{2}V^{\prime}(|\mathbf{r}%
_{1}-\mathbf{r}_{2}|)\frac{(\mathbf{\dot{r}}_{1}-\mathbf{\dot{r}}_{2}%
)}{|\mathbf{r}_{1}-\mathbf{r}_{2}|}\cdot(\mathbf{r}_{1}-\mathbf{r}%
_{2})(\mathbf{r}_{1}+\mathbf{r}_{2})
\end{align}
The momenta can be obtained from the Lagrangian in Eq. (28) as%
\begin{align}
\mathbf{p}_{1}  &  =\frac{\partial L}{\partial\mathbf{\dot{r}}_{1}}%
=m_{1}\mathbf{\dot{r}}_{1}\left(  1-\frac{\mathbf{\dot{r}}_{1}^{2}}{c^{2}%
}\right)  ^{-1/2}+\frac{1}{2}V(|\mathbf{r}_{1}-\mathbf{r}_{2}|)\frac
{\mathbf{\dot{r}}_{2}}{c^{2}}\nonumber\\
&  -\frac{1}{2}V^{\prime}(|\mathbf{r}_{1}-\mathbf{r}_{2}|)\frac{(\mathbf{r}%
_{1}-\mathbf{r}_{2})}{\left\vert \mathbf{r}_{1}-\mathbf{r}_{2}\right\vert
}\frac{\mathbf{\dot{r}}_{2}}{c^{2}}\cdot(\mathbf{r}_{1}-\mathbf{r}_{2})
\end{align}%
\begin{align}
\mathbf{p}_{2}  &  =\frac{\partial L}{\partial\mathbf{\dot{r}}_{2}}%
=m_{2}\mathbf{\dot{r}}_{2}\left(  1-\frac{\mathbf{\dot{r}}_{2}^{2}}{c^{2}%
}\right)  ^{-1/2}+\frac{1}{2}V(|\mathbf{r}_{1}-\mathbf{r}_{2}|)\frac
{\mathbf{\dot{r}}_{1}}{c^{2}}\nonumber\\
&  -\frac{1}{2}V^{\prime}(|\mathbf{r}_{1}-\mathbf{r}_{2}|)\frac{(\mathbf{r}%
_{1}-\mathbf{r}_{2})}{\left\vert \mathbf{r}_{1}-\mathbf{r}_{2}\right\vert
}\frac{\mathbf{\dot{r}}_{1}}{c^{2}}\cdot(\mathbf{r}_{1}-\mathbf{r}_{2})
\end{align}
giving total linear momentum%
\begin{align}
\mathbf{P}  &  \mathbf{=}m_{1}\mathbf{\dot{r}}_{1}\left(  1-\frac
{\mathbf{\dot{r}}_{1}^{2}}{c^{2}}\right)  ^{-1/2}+m_{2}\mathbf{\dot{r}}%
_{2}\left(  1-\frac{\mathbf{\dot{r}}_{2}^{2}}{c^{2}}\right)  ^{-1/2}%
\nonumber\\
&  +\frac{1}{2}V(|\mathbf{r}_{1}-\mathbf{r}_{2}|)\frac{\mathbf{\dot{r}}_{1}%
}{c^{2}}+\frac{1}{2}V(|\mathbf{r}_{1}-\mathbf{r}_{2}|)\frac{\mathbf{\dot{r}%
}_{2}}{c^{2}}\nonumber\\
&  -\frac{1}{2}V^{\prime}(|\mathbf{r}_{1}-\mathbf{r}_{2}|)\frac{(\mathbf{r}%
_{1}-\mathbf{r}_{2})}{\left\vert \mathbf{r}_{1}-\mathbf{r}_{2}\right\vert
}\frac{\mathbf{\dot{r}}_{1}}{c^{2}}\cdot(\mathbf{r}_{1}-\mathbf{r}_{2}%
)-\frac{1}{2}V^{\prime}(|\mathbf{r}_{1}-\mathbf{r}_{2}|)\frac{(\mathbf{r}%
_{1}-\mathbf{r}_{2})}{\left\vert \mathbf{r}_{1}-\mathbf{r}_{2}\right\vert
}\frac{\mathbf{\dot{r}}_{2}}{c^{2}}\cdot(\mathbf{r}_{1}-\mathbf{r}_{2})
\end{align}
Comparing Eqs. (30) and (36) after reorganizing a few terms and expanding any
factors of $\gamma=(1-v^{2}/c^{2})^{-1/2}$, we find that indeed Eq. (4) holds.
\ The system of Eq. (28) is indeed Lorentz invariant through order
$v^{2}/c^{2}.$

We notice from Eqs. (34)-(36) that momentum is now no longer connected
exclusively with restmass times velocity, as is required by the fourth
conservation law for Galilean invariance. \ The center of restmass for the
system of Eq. (28) no longer moves with constant velocity. \ Now linear
momentum is associated with the parameters of the potential energy, and the
system is being transformed toward a field-theory point of view.

\subsection{Velocity-Dependent and Acceleration-Dependent Forces in
Lorentz-Invariant Systems Through Order $v^{2}/c^{2}$}

The Lagrange equations of motion follow from the Lagrangian in Eq. (28); for
the particle at $\mathbf{r}_{1}$, the equation takes the form
\begin{align}
0  &  =\frac{d}{dt}\left(  \frac{m_{1}\mathbf{\dot{r}}_{1}}{(1-\mathbf{\dot
{r}}_{1}^{2}/c^{2})^{1/2}}+\frac{1}{2}V(|\mathbf{r}_{1}-\mathbf{r}_{2}%
|)\frac{\mathbf{\dot{r}}_{2}}{c^{2}}-\frac{1}{2}V^{\prime}(|\mathbf{r}%
_{1}-\mathbf{r}_{2}|)\frac{(\mathbf{r}_{1}-\mathbf{r}_{2})}{\left\vert
\mathbf{r}_{1}-\mathbf{r}_{2}\right\vert }\frac{\mathbf{\dot{r}}_{2}}{c^{2}%
}\cdot(\mathbf{r}_{1}-\mathbf{r}_{2})\right) \nonumber\\
&  -\frac{\mathbf{r}_{1}-\mathbf{r}_{2}}{\left\vert \mathbf{r}_{1}%
-\mathbf{r}_{2}\right\vert }V^{\prime}(|\mathbf{r}_{1}-\mathbf{r}_{2}|)\left(
-1+\frac{\mathbf{\dot{r}}_{1}\cdot\mathbf{\dot{r}}_{2}}{2c^{2}})+\frac
{\mathbf{\dot{r}}_{1}\cdot(\mathbf{r}_{1}-\mathbf{r}_{2})\mathbf{\dot{r}}%
_{2}\cdot(\mathbf{r}_{1}-\mathbf{r}_{2})}{2c^{2}\left\vert \mathbf{r}%
_{1}-\mathbf{r}_{2}\right\vert ^{2}}\right)  \text{ \ }\nonumber\\
&  +\frac{\mathbf{r}_{1}-\mathbf{r}_{2}}{\left\vert \mathbf{r}_{1}%
-\mathbf{r}_{2}\right\vert }V^{^{\prime\prime}}(|\mathbf{r}_{1}-\mathbf{r}%
_{2}|)\frac{\mathbf{\dot{r}}_{1}\cdot(\mathbf{r}_{1}-\mathbf{r}_{2}%
)\mathbf{\dot{r}}_{2}\cdot(\mathbf{r}_{1}-\mathbf{r}_{2})}{2c^{2}\left\vert
\mathbf{r}_{1}-\mathbf{r}_{2}\right\vert ^{2}}\nonumber\\
&  +V^{\prime}(|\mathbf{r}_{1}-\mathbf{r}_{2}|)\left(  \frac{\mathbf{\dot{r}%
}_{1}\cdot(\mathbf{r}_{1}-\mathbf{r}_{2})\mathbf{\dot{r}}_{2}+\mathbf{\dot{r}%
}_{2}\cdot(\mathbf{r}_{1}-\mathbf{r}_{2})\mathbf{\dot{r}}_{1}}{2c^{2}%
\left\vert \mathbf{r}_{1}-\mathbf{r}_{2}\right\vert }\right)
\end{align}
The equations of motion can be rewritten as forces acting on the particles to
change the mechanical momentum. \ For the particle at $\mathbf{r}_{1}$, this
becomes%
\begin{align}
\frac{d}{dt}\left(  \frac{m\mathbf{\dot{r}}_{1}}{(1-\mathbf{\dot{r}}_{1}%
^{2}/c^{2})^{1/2}}\right)   &  =-V^{\prime}(|\mathbf{r}_{1}-\mathbf{r}%
_{2}|)\frac{\mathbf{r}_{1}-\mathbf{r}_{2}}{\left\vert \mathbf{r}%
_{1}-\mathbf{r}_{2}\right\vert }\left[  1+\frac{1}{2}\left(  \frac
{\mathbf{\dot{r}}_{2}}{c}\right)  ^{2}\right] \nonumber\\
&  -\frac{\mathbf{r}_{1}-\mathbf{r}_{2}}{2c^{2}\left\vert \mathbf{r}%
_{1}-\mathbf{r}_{2}\right\vert }\left(  \frac{V^{^{\prime\prime}}%
(|\mathbf{r}_{1}-\mathbf{r}_{2}|)}{\left\vert \mathbf{r}_{1}-\mathbf{r}%
_{2}\right\vert }-\frac{V^{\prime}(|\mathbf{r}_{1}-\mathbf{r}_{2}%
|)}{\left\vert \mathbf{r}_{1}-\mathbf{r}_{2}\right\vert ^{2}}\right)  \left[
\mathbf{\dot{r}}_{2}\cdot(\mathbf{r}_{1}-\mathbf{r}_{2})\right]
^{2}\nonumber\\
&  -\frac{1}{2c^{2}}\left(  V(|\mathbf{r}_{1}-\mathbf{r}_{2}|)\mathbf{\ddot
{r}}_{2}-V^{\prime}(|\mathbf{r}_{1}-\mathbf{r}_{2}|)\frac{[\mathbf{\ddot{r}%
}_{2}\cdot(\mathbf{r}_{1}-\mathbf{r}_{2})](\mathbf{r}_{1}-\mathbf{r}_{2}%
)}{|\mathbf{r}_{1}-\mathbf{r}_{2}|^{2}}\right) \nonumber\\
&  -\frac{\mathbf{\dot{r}}_{1}}{c}\times\left(  \frac{\mathbf{\dot{r}}_{2}}%
{c}\times\frac{\mathbf{r}_{1}-\mathbf{r}_{2}}{\left\vert \mathbf{r}%
_{1}-\mathbf{r}_{2}\right\vert }V^{\prime}(|\mathbf{r}_{1}-\mathbf{r}%
_{2}|)\right)
\end{align}
We notice that the force on the first particle involves not only the force
arising from the original nonrelativistic potential function, but also forces
depending upon the velocities of both particles and upon the acceleration of
the other particle. These forces were not part of the original nonrelativistic
theory. \ Such forces are absent from the accounts in the mechanics
textbooks\cite{Goldstein3} and from the articles which treat "relativistic"
motion for a single particle. The single particle appearing in the Lagrangian
of these treatments actually produces velocity-dependent and
acceleration-dependent forces back on the prescribed sources whose momentum
and energy are never discussed.

The most famous Lagrangian which is Lorentz invariant through $v^{2}/c^{2}$ is
that obtained from the Coulomb potential $V(|\mathbf{r}_{1}-\mathbf{r}%
_{2}|)=q_{1}q_{2}/|\mathbf{r}_{1}-\mathbf{r}_{2}|.$ \ In this case the
Lagrangian of Eq. (28) becomes
\begin{align}
L(\mathbf{r}_{1},\mathbf{r}_{2},\mathbf{\dot{r}}_{1},\mathbf{\dot{r}}_{2}) &
=-m_{1}c^{2}(1-\mathbf{\dot{r}}_{1}^{2}/c^{2})^{1/2}-m_{2}c^{2}(1-\mathbf{\dot
{r}}_{2}^{2}/c^{2})^{1/2}-\frac{q_{1}q_{2}}{|\mathbf{r}_{1}-\mathbf{r}_{2}%
|}\nonumber\\
&  +\frac{1}{2}\frac{q_{1}q_{2}}{|\mathbf{r}_{1}-\mathbf{r}_{2}|}%
\frac{\mathbf{\dot{r}}_{1}\cdot\mathbf{\dot{r}}_{2}}{c^{2}}+\frac{1}{2}%
\frac{q_{1}q_{2}}{|\mathbf{r}_{1}-\mathbf{r}_{2}|}\frac{\mathbf{\dot{r}}%
_{1}\cdot(\mathbf{r}_{1}-\mathbf{r}_{2})\mathbf{\dot{r}}_{2}\cdot
(\mathbf{r}_{1}-\mathbf{r}_{2})}{c^{2}\left\vert \mathbf{r}_{1}-\mathbf{r}%
_{2}\right\vert ^{2}}%
\end{align}
If in Eq. (39) we expand the free-particle expressions $-mc^{2}(1-\mathbf{\dot
{r}}^{2}/c^{2})^{1/2}$ through second order in $v/c,$ then this becomes the
Darwin Lagrangian which sometime appears in electromagnetism textbooks as an
approximation to the interaction of charged particles.\cite{Jack} \ The
approximation is an accurate description of the classical electromagnetic
interaction between charged particles through second order in $v/c$ for small
separations between the particles. \ The Lagrangian equation of motion
following from Eq.(39) becomes (for the particle at position $\mathbf{r}_{1}$)%
\begin{align}
\frac{d}{dt}\left(  \frac{m_{1}\mathbf{\dot{r}}_{1}}{(1-\mathbf{\dot{r}}%
_{1}^{2}/c^{2})^{1/2}}\right)   &  =q_{1}[q_{2}\frac{(\mathbf{r}%
_{1}-\mathbf{r}_{2})}{|\mathbf{r}_{1}-\mathbf{r}_{2}|^{3}}\left\{  1+\frac
{1}{2}\left(  \frac{\mathbf{\dot{r}}_{2}}{c}\right)  ^{2}-\frac{3}{2}\left(
\frac{(\mathbf{r}_{1}-\mathbf{r}_{2})\cdot\mathbf{\dot{r}}_{2}}{c|\mathbf{r}%
_{1}-\mathbf{r}_{2}|}\right)  ^{2}\right\}  \nonumber\\
&  -\frac{q_{2}}{2c^{2}}\left(  \mathbf{\ddot{r}}_{2}\mathbf{+}\frac
{[\mathbf{\ddot{r}}_{2}\cdot(\mathbf{r}_{1}-\mathbf{r}_{2})](\mathbf{r}%
_{1}-\mathbf{r}_{2})}{|\mathbf{r}_{1}-\mathbf{r}_{2}|^{2}}\right)
]\nonumber\\
&  +q_{1}\frac{\mathbf{\dot{r}}_{1}}{c}\times\left[  q_{2}\frac{\mathbf{\dot
{r}}_{2}}{c}\times\frac{(\mathbf{r}_{1}-\mathbf{r}_{2})}{|\mathbf{r}%
_{1}-\mathbf{r}_{2}|^{3}}\right]
\end{align}
where we have rewritten the Lagrangian equation in the form $d\mathbf{p}%
_{1}/dt=q_{1}\mathbf{E+}q_{1}(\mathbf{\dot{r}}_{1}/c)\times\mathbf{B}$ with
$\mathbf{p}_{1}$ the mechanical particle momentum.\cite{PA} \ The velocity-
and acceleration-dependent forces in Eq.(40) correspond to fields arising from
electromagnetic induction. \ In the textbooks, electromagnetic induction is
always treated without reference to any charged particles which may be
producing the induction fields, a very different point of view from that which
follows from the Darwin Lagrangian.

Contemporary physics regards special relativity as a metatheory to which
(locally) all theories describing nature should conform. \ Thus in
nonrelativistic classical mechanics, there is the unspoken implication that
the nonrelativistic interaction between point particles at positions
$\mathbf{r}_{1}$ and $\mathbf{r}_{2}$ under a general potential $V(|\mathbf{r}%
_{1}-\mathbf{r}_{2}|)$ is the small-velocity limit of some fully relativistic
theory of interacting point particles which might occur in nature. \ However,
the use of a general potential can be misleading for both students and
researchers. Here we demonstrate that an arbitrary nonrelativistic potential
function can indeed be extended to a Lagrangian which is Lorentz-invariant
through order $v^{2}/c^{2};$ however, the extension requires the introduction
of velocity-dependent and acceleration-dependent forces which go unmentioned
in the mechanics textbooks.

\section{Discussion}

In this paper, we have tried to give simple examples which illustrate the
conservation laws of Lorentz-invariant systems and which suggest the reason
for the "no-interaction theorem" of relativistic mechanics beyond point binary
collisions. \ The introduction of position-dependent potential energy places a
burden on the fourth relativistic conservation law requiring the constant
velocity of the center of energy. \ In general this burden is so restrictive
that there can be no interaction through a potential without an accompanying
field which carries both energy and linear momentum.

The treatments of relativistic mechanics in textbooks can be misleading. \ In
mechanics text books, we often find discussions of single-particle motion in
potentials.\cite{Goldstein3} \ The most famous (and appropriate example) is
the motion of an particle in a Coulomb potential. \ This can be regarded as an
approximation to the fully relativistic two-charged-particle electromagnetic
interaction where the nucleus is much more massive than the orbiting electron,
and so the classical electrodynamic analysis agrees with the one-particle
mechanical analysis in the approximation which ignores emission and/or
absorption by the electromagnetic radiation field. \ However, there is also a
mechanics text book which discusses a particle with relativistic momentum in a
harmonic oscillator potential, with no mention of field
theory.\cite{Goldstein2} \ An uninformed student might easily assume that this
is an example of a relativistic interaction between two particles, one of
which is much more massive than the other, which are interacting through the
harmonic oscillator potential. \ We have emphasized that the no-interaction
theorem indicates that such a purely mechanical relativistic interaction is
not possible. \ 

Classical electromagnetism is a relativistic theory which was developed during
the nineteenth century before the ideas of special relativity. \ Indeed,
special relativity arose at the beginning of the twentieth century as a
response to the conflict of electromagnetism with nonrelativistic mechanics.
\ Around the same time, quantum mechanics was introduced in response to the
mismatch between electromagnetic radiation equilibrium (blackbody radiation)
and classical statistical mechanics (which is based on nonrelativistic
mechanics). \ Although quantum theory and special relativity have gone on to
enormous successes, they have left behind a number of unresolved questions
within classical physics. \ For example, the blackbody radiation problem has
never been solved within relativistic classical physics.\cite{bl} There have
been discussions of classical radiation equilibrium using nonrelativistic
mechanical scatterers and even one calculation of a scattering particle using
relativistic mechanical momentum in a general class of non-Coulomb
potentials.\cite{Blanco} \ However, there has never been a treatment of
scattering by a relativistic particle in the Coulomb potential of classical
electrodynamics, despite the fact that the Coulomb potential has all the
qualitative aspects which might allow classical radiation equilibrium at a
spectrum with finite thermal energy.

We conclude that misconceptions regarding potentials which can be regarded as
approximations to relativistic systems are relevant for treatments in
mechanics textbooks and perhaps also for the description of nature within
classical theory.

Acknowledgement

I wish to thank Professor V. Parameswaran Nair for a discussion regarding the
no-interaction theorem of Currie, Jordan, and Sudarshan, and also to thank
Professor Hanno Ess\'{e}n for bringing to my attention the work listed in
reference 9.

\end{document}